\documentclass[prx,twocolumn, amsmath,amssymb,
reprint,%
author-year,%
author-numerical,%
dvipdfmx
]{revtex4}

\usepackage{graphicx}
\usepackage{bm}
\usepackage{color}

\begin{document}


\title{Hypothetical FrAu: An outlier in the B2 compounds}

\author{Shota Ono}
\email{shota\_o@gifu-u.ac.jp}
\affiliation{Department of Electrical, Electronic and Computer Engineering, Gifu University, Gifu 501-1193, Japan}

\begin{abstract}
The ordered alloys of alkali metals (Rb and Cs) and gold (Au) have the B2 (CsCl-type) structure and show a semiconducting property, irrespective to the metallic constituents. Francium (Fr) is classed as an alkali metal and is expected to form the B2 structure with Au. However, it is difficult to synthesize such a compound experimentally due to a half-life of a few ten minutes in the Fr atom. In this paper, by using the first-principles method, we study the structural and electronic properties of FrAu in the B2 structure. The FrAu has a relatively large lattice constant and a relatively small bulk modulus among 310 B2 compounds. The profiles of the electron and phonon band structures of the FrAu are quite similar to those of the CsAu. We predict that the FrAu is included to one of the ionic compounds as well as the RbAu and CsAu.
\end{abstract}

\maketitle

{\it Introduction.---}Francium (Fr) was discovered by Marguerite Perey in 1939. It has the atomic number of 87 and has a half-life of only 22 minutes, so that it is difficult to study the physical and chemical properties experimentally in detail. In a quantum mechanical point of view, the Fr atom has one $7s$ electron and is classed as an alkali metal. The chemical property is thus expected to be similar to those of other alkali metals such as rubidium (Rb) and cesium (Cs). The Pauling electronegativity of alkali metals decreases monotonically with the period in the periodic table \cite{pauling}. The electronegativity of Fr is expected to be smaller than that of Cs, while the trend of the magnitude may depend on the definition of the electronegativity \cite{webelements}. It has not been understood well what structural and electronic properties the Fr-based compounds have. 

The trends of electronic properties of $A$Au in the B2 (CsCl-type) structure ($A=$ Li, Na, K, Rb, and Cs) are interesting because the LiAu, NaAu, and KAu are metal, whereas the RbAu and CsAu are semiconductor \cite{spicer}. The optical bandgap of CsAu was reported to be 2.6 eV \cite{spicer}, while the indirect bandgap was predicted to be 0.3 eV for RbAu \cite{RbAu} and about 1.0 eV for CsAu \cite{koenig,CsAu} within the density-functional theory (DFT) calculations. The molecular dynamics simulations have predicted that the liquid CsAu has a frequency gap between the longitudinal optical (LO) and transverse optical (TO) branches in the limit of long wavelength, also showing a non-metallic character \cite{bryk}. The semiconducting property of the RbAu and CsAu is due to formation of bonds with ionic character, which originates from the large difference of the electronegativity between alkali metals and gold \cite{koenig}. In this sense, the FrAu in the B2 structure would have a bandgap whose magnitude is comparable to that of CsAu. 

The B2 structure is commonly observed for many compounds in their thermodynamic phase diagram \cite{sluiter,kolli}. In recent years, several properties of the B2 compounds have been studied within the DFT calculations: For example, the stability trend in CuZn, AgZn, and AuZn \cite{alsalmi}, the disordered phase in FeAl \cite{FeAl}, the finite temperature effect in HgMg \cite{HgMg}, and the metastability in the Cu-based compounds \cite{ono2021}. In this paper, we study the trends of structural and electronic properties of $A$Au in the B2 structure ($A=$ Rb, Cs, and Fr). We demonstrate that the FrAu has a relatively large lattice parameter and a relatively small bulk modulus among 310 B2 compounds. The profile of the electron and phonon band structures are similar between the CsAu and FrAu. While the FrAu is a hypothetical compound due to a short half life, it can be included to one of the ionic compounds as well as the RbAu and CsAu. 


\begin{figure*}
\center
\includegraphics[scale=0.6]{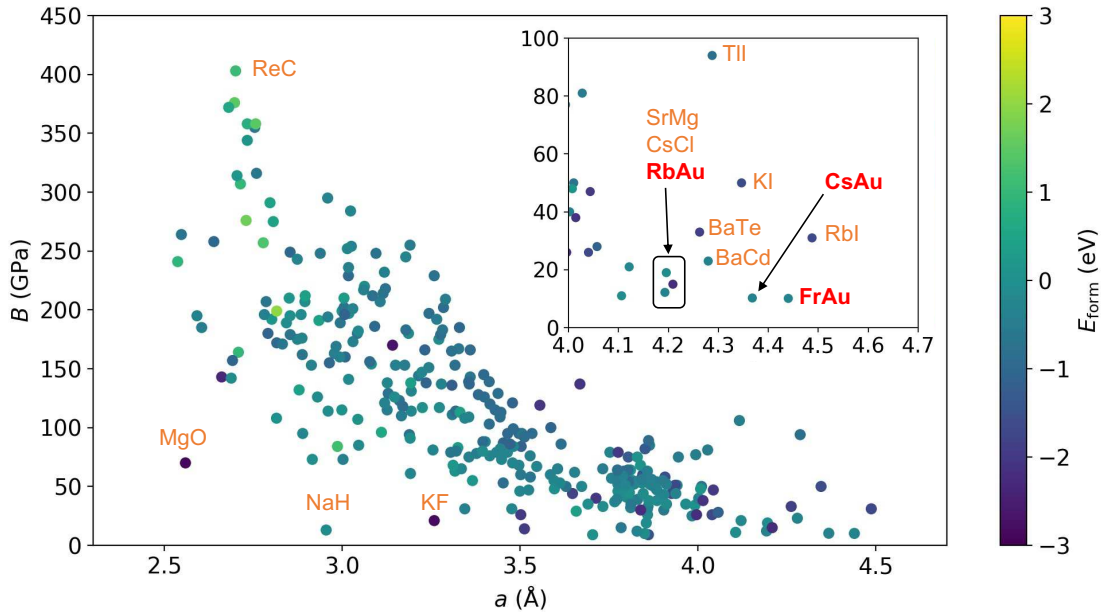}
\caption{Distribution of $a$, $B$, and $E_{\rm form}$ for 310 B2 compounds. Except for RbAu, CsAu, and FrAu (red), the values were extracted from the Materials Project database \cite{MP} by using the pymatgen \cite{pymatgen}. Inset: a magnified view for $a \in [4.0,4.7]$ \AA \ and $B\in [0,100]$ GPa. } \label{fig1} 
\end{figure*}

\begin{figure*}
\center
\includegraphics[scale=0.55]{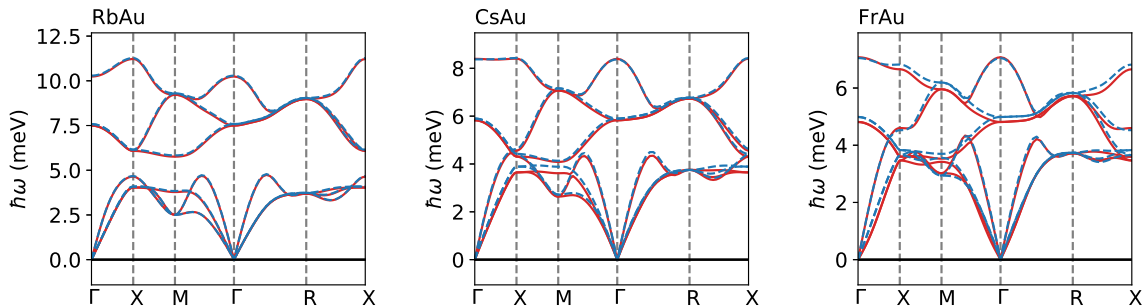}
\caption{The phonon band structure of RbAu, CsAu, and FrAu without SOC (solid red) and with SOC (dashed blue).} \label{fig2} 
\end{figure*}

\begin{figure}
\center
\includegraphics[scale=0.5]{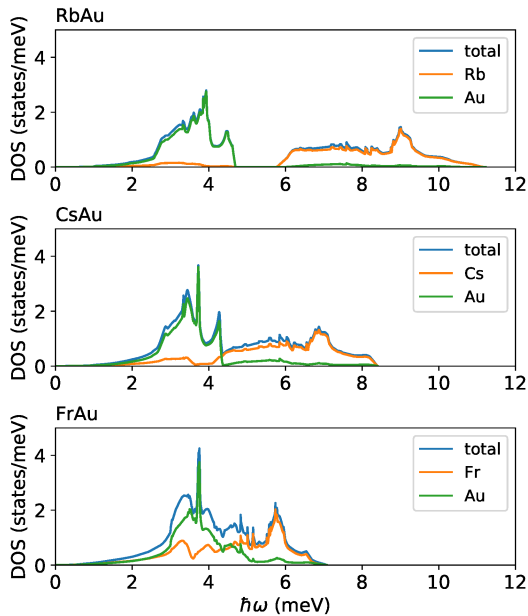}
\caption{The phonon DOS of RbAu, CsAu, and FrAu.} \label{fig3} 
\end{figure}

\begin{figure*}
\center
\includegraphics[scale=0.55]{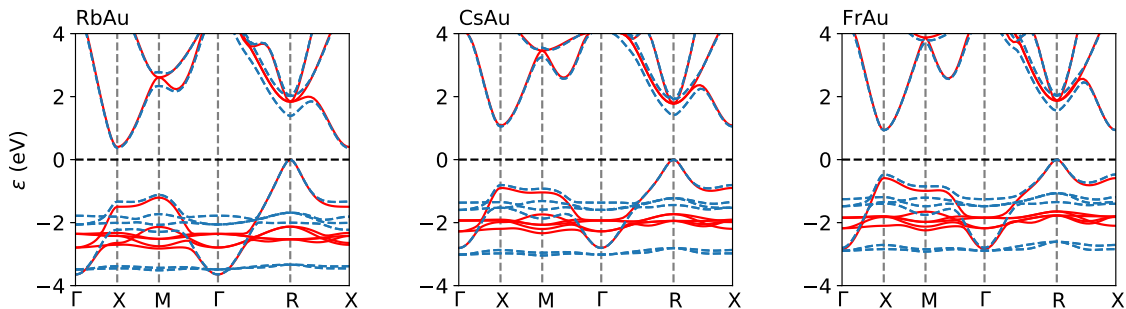}
\caption{The electron band structure of RbAu, CsAu, and FrAu without SOC (solid red) and with SOC (dashed blue).} \label{fig4} 
\end{figure*}

\begin{figure}
\center
\includegraphics[scale=0.5]{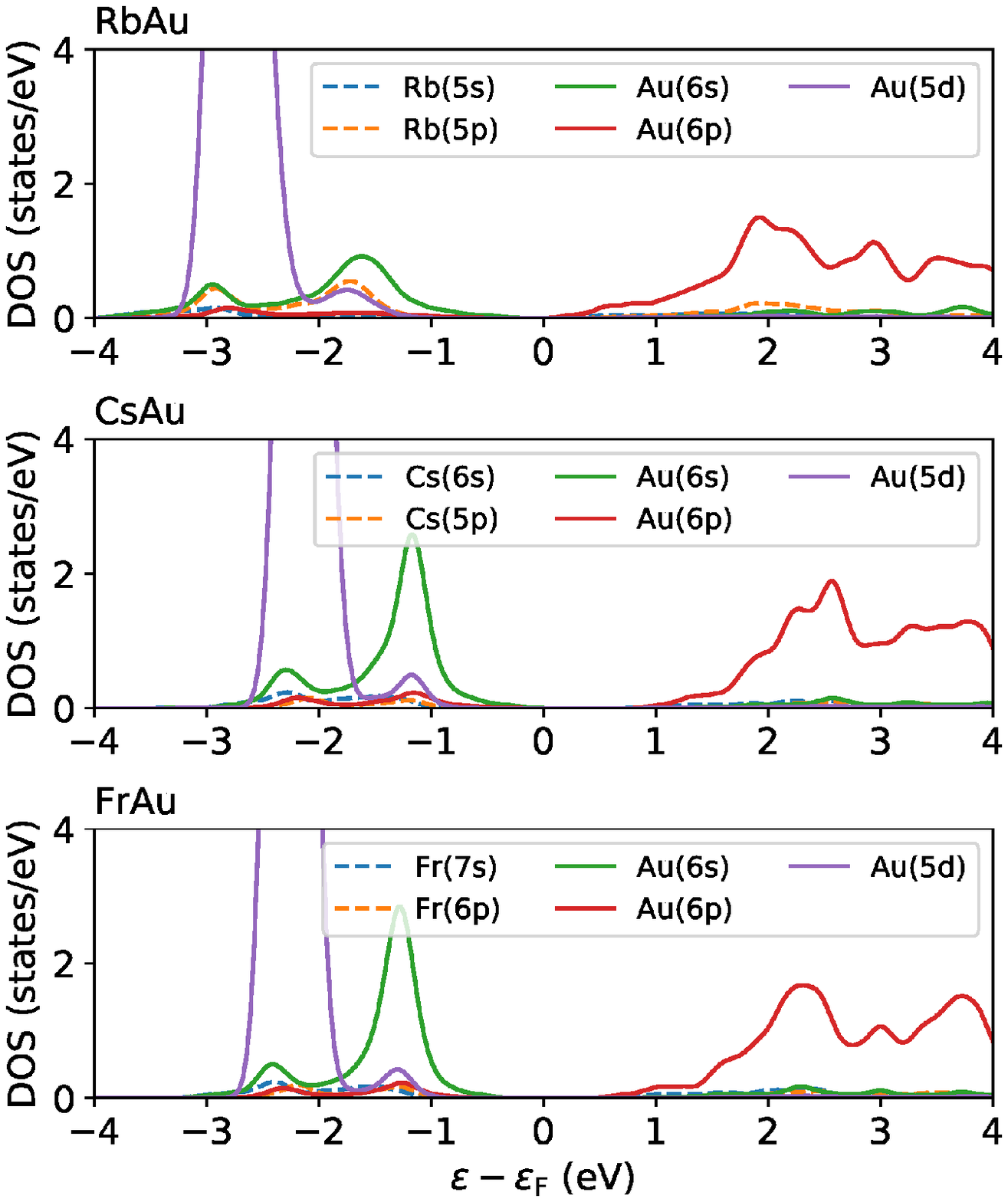}
\caption{The electron DOS of RbAu, CsAu, and FrAu.} \label{fig5} 
\end{figure}


{\it Methods.---}We performed DFT calculations for B2 $A$Au ($A=$ Rb, Cs, and Fr) by using \texttt{Quantum ESPRESSO} (\texttt{QE}) \cite{qe}. We used the Perdew-Burke-Ernzerhof (PBE) \cite{pbe} functionals of the generalized gradient approximation for the exchange-correlation energy and used the ultrasoft pseudopotentials provided in \texttt{pslibrary.1.0.0} \cite{dalcorso}. The effect of the spin-orbit coupling (SOC) was treated within the scheme in Refs.~\cite{soc1,soc2}. The cutoff energies for the wavefunction and the charge density are 80 Ry and 800 Ry, respectively, and 14$\times$14$\times$14 $k$ grid was used in the self-consistent field (scf) calculations \cite{MK}. The mass of the Rb, Cs, Fr, and Au atoms was set to be 85.467, 132.905, 223, and 196.966 a.u., respectively. 

The lattice parameter $a$ was optimized by using the bisection method, where the total energy was converged within $10^{-7}$ Ry. Then, we calculated the bulk modulus $B$, the formation energy $E_{\rm form}$, the electron and phonon band structures of $A$Au. The magnitude of $B$ is defined as
\begin{eqnarray}
 B &=& V\left(\frac{\partial^2 E}{\partial V^2} \right)_0 
 = \frac{1}{9a}\left(\frac{\partial^2 E}{\partial a^2} \right)_0,  
\end{eqnarray}
where $E$ and $V$ are the total energy and the volume of the unit cell, respectively. The derivative is taken at the volume of the relaxed structure. This was evaluated by applying the deformation of $\pm \delta a$ and $\pm 2\delta a$, where $\delta a$ was set to be $0.005a$. The $E_{\rm form}$ of $A$Au is defined as
\begin{eqnarray}
 E_{\rm form}(A{\rm Au}) = \varepsilon_{\rm B2}(A{\rm Au}) - \varepsilon_{\rm bcc}(A) - \varepsilon_{\rm fcc}({\rm Au}),
\end{eqnarray} 
where $\varepsilon_{j}(X)$ is the total energy of the material $X$ in the structure $j$. For the simple metals in the body-centered cubic (bcc) and the face-centered cubic (fcc) structures, the size of the $k$ grid was increased to 24$\times$24$\times$24 and an improved tetrahedron method \cite{tetra_opt} was used. For the electron band structure calculations, the improved tetrahedron method was also adapted for creating the charge density. The phonon dispersions of $A$Au were calculated within the density-functional perturbation theory (DFPT) \cite{dfpt} implemented in \texttt{QE} \cite{qe}. 4$\times$4$\times$4 $q$ grid ($10 \ q$ points in the Brillouin zone) was used. 

Note that the cohesive energy $E_{\rm coh}$ of the Fr in the simple cubic (sc), the fcc, the bcc, and the hexagonal close packed (hcp) structures was estimated to be 0.625, 0.700, 0.701, and 0.704 eV per atom, respectively, where the PBE without the SOC was used. In other alkali metals \cite{wang}, the hcp structure has also been predicted to be more stable than the bcc structure. However, the alkali metals have the bcc structure in the experimental phase diagram. We thus assume that the Fr has the bcc structure when we estimate the $E_{\rm form}$ of $A$Au. We also confirmed that the FrAu in the B2 structure is more stable than that in the B1 (NaCl-type) structure by 1 eV per atom. 


When we used the Perdew-Zunger functional \cite{pz}, we found that the RbAu and CsAu have imaginary phonon frequencies at the M point, i.e., the B2 structure is unstable, while the FrAu is dynamically stable. This is not consistent with the experimental synthesis of RbAu and CsAu in the B2 structure \cite{spicer,CsAuexp}. We will show the results calculated by using the PBE functional below. 

\begin{table*}
\begin{center}
\caption{The lattice constant $a$ (\AA), the bulk modulus $B$ (GPa), the formation energy $E_{\rm form}$ (eV), and the dielectric constant $\varepsilon_{\rm infty}$ of $A$Au. The figures in the parenthesis indicate the calculated values with the SOC included. The experimental values of $a$ are extracted from Ref.~\cite{CsAuexp}. }
{
\begin{tabular}{lccccccccc}\hline\hline
 $A$  \hspace{5mm} & $a$ &  & exp. & \hspace{5mm} $B$ &  & \hspace{5mm}  $E_{\rm form}$ & & \hspace{5mm}  $\varepsilon_{\infty}$ & \\  \hline
Rb & 4.193 & (4.186) & 4.098 & \hspace{5mm}  12.18 & (12.37) & \hspace{5mm} $-0.283$ & ($-1.313$) & \hspace{5mm}  5.273 & (7.287) \\

Cs & 4.368 & (4.359) & 4.258 & \hspace{5mm}  10.24  & (10.38) & \hspace{5mm} $-0.334$ & ($-1.300$) & \hspace{5mm}  4.966 & (7.009)  \\

Fr & 4.440 & (4.408) & -          & \hspace{5mm}  10.10  & (10.46) &\hspace{5mm}  $-0.259$ & ($-1.254$) & \hspace{5mm}  5.304 & (7.963)  \\

\hline\hline
\end{tabular}
}
\label{table_all}
\end{center}
\end{table*}

{\it Structural property.---}We first provide a correlation between the $B$ and the $a$ for the compounds in the B2 structure (see Fig.~\ref{fig1}). Except for the RbAu, CsAu, and FrAu, the plotted data were extracted from the Materials Project (MP) database \cite{MP} by using the pymatgen program \cite{pymatgen}, where the materials with the spacegroup of $Pm\bar{3}m$ and the number of sites in the unit cell of $2$ were assumed. When the $a$ increases, the $B$ tends to decrease. The RbAu, CsAu, and FrAu follow this trend in the B2 compounds. However, they can be regarded as an outlier with extremely large $a$, as shown in the inset of Fig.~\ref{fig1}. The values of $a$ (\AA) and $B$ (GPa) of $A$Au are listed in Table \ref{table_all}.


The value of $a$ increases monotonically as one goes from Rb to Fr. The $a$ for $A=$ Fr is larger than that for $A=$ Cs by only less than 0.1 \AA. An increase in $a$ reflects the repulsive forces between ions, that is, the larger the atomic number of $A$, the larger the direct Coulomb interaction forces between adjacent atoms ($A$-Au and $A$-$A$). The slight difference of $a$ between CsAu and FrAu will be attributed to an attractive force mediated by the ion-electron-ion interaction: the $7s$ electron is largely extended around the Fr atom, compared to the $6s$ electron in the Cs atom, forming a strong bonding between atoms. When the SOC is included, the magnitude of $a$ decreases less than one percent.


The value of $B$ decreases as one goes from Rb to Fr, while the difference between the Cs and Fr is only 0.14 GPa. This is a similar tendency for the $a$, where a small difference of $a$ between the CsAu and FrAu was identified. With the SOC included, the magnitude of $B$ increases, yielding an inequality of $B({\rm CsAu})<B({\rm FrAu})$. 


The $E_{\rm form}$s of $A$Au are also listed in Table \ref{table_all}. The FrAu has the highest value of $E_{\rm form}$, indicating that this is energetically unstable among the $A$Au. On the other hand, the CsAu has the lowest value of $E_{\rm form}$. As shown in Fig.~\ref{fig1}, the compounds with a small negative $E_{\rm form}$ are distributed over a wide range of $a$. These can have various types of bonding: the ionic, covalent, metallic, and their mixed characters. It will be desirable to develop a theory for understanding the energetics in a unified way.  

Note that the MP database can include the theoretically predicted materials as well as the experimentally reported materials. This implies that some of them are unstable at ambient condition. For example, the B2 MgO with $a_{\rm lat}=2.66$ \AA, $B=143$ GPa, and $E_{\rm form}=-2.30$ eV/atom will be transformed into the B1 structure and the B2 GdIn with $a_{\rm lat}=3.83$ \AA, $B=376$ GPa, and $E_{\rm form}=-0.48$ eV/atom (hidden in the inset of Fig.~\ref{fig1}) will be transformed into the L1$_0$ (CuAu-type) structure. 

{\it Dynamical stability.---}To study the dynamical stability of $A$Au, we show the phonon band structure of RbAu, CsAu, and FrAu within PBE in Fig.~\ref{fig2}. All frequencies within the Brillouin zone are positive, showing that these compounds in the B2 structure are dynamically stable. Also the LO-TO splitting at the $\Gamma$ point is observed, which is due to finite values of the dielectric constant $\varepsilon_{\infty}$ (calculated by using \texttt{QE}) listed in Table \ref{table_all}. The non-metallic character of the liquid CsAu has also been identified by the LO-TO splitting \cite{bryk}. The effect of the SOC is negligibly small for RbAu, while it can cause a small phonon hardening in the acoustic and optical phonon branches for CsAu and FrAu. 

To understand the contribution to the vibrational spectra from each atom (i.e., the alkali metal $A$ and Au), the atom-projected density-of-states (DOS) is shown in Fig.~\ref{fig3}. For RbAu, the mass of the Rb atom is small enough to open a frequency gap between 5 to 6 meV. This is equivalent to the fact that the high frequency DOS is mainly dominated by the Rb atom vibrations. When the Rb atom is replaced with the Cs atom, the frequency gap closes and the Au vibration contributes to the optical modes. When the Cs atom is replaced with the Fr atom that is heavier than the Au atom, the DOS peak around 6 meV is still dominated by the Fr atom vibrations. 

{\it Electronic property.---}Figure \ref{fig4} shows the electron band structure of RbAu, CsAu, and FrAu. The indirect bandgap is observed, where the valence band top and the conduction band bottom are located at R and X point, respectively. The magnitude of the indirect gandgap $E_g$ is estimated to be 0.40, 1.08, and 0.93 eV for RbAu, CsAu, and FrAu, respectively. These values are not changed significantly when the SOC is included. The impact of the SOC is observed in the valence band: the flat bands around $\varepsilon\simeq -2$ eV without the SOC split into two bands that are separated by 1.5 eV. 

Figure \ref{fig5} shows the electron DOS decomposed into the atomic orbitals of $A$Au by using the PBE without the SOC. The Au $5d$ orbital produces a high DOS peak (the peak height is about 25 states/eV for all cases) below the Fermi level by around a few eV, explaining the flat bands shown in Fig.~\ref{fig4}. The Au $6s$ and $6p$ orbitals contribute to the valence and conduction bands, respectively, forming parabolic bands shown in Fig.~\ref{fig4}. The alkali atoms have a minor contribution to the electron bands around the Fermi level. The profile of the DOS between the CsAu and FrAu is also similar. 

The ionic character of the $A$Au can be identified by studying the charge transfer between the alkali metal $A$ and the Au. We calculated the difference of the L\"{o}wdin charge \cite{lowdin} for the atom $X \ (=A \ {\rm or \ Au})$ defined as 
\begin{eqnarray}
 \Delta N_{A{\rm Au}}(X) = N_{A{\rm Au}}(X)- N_{\rm atom}(X),
\end{eqnarray}
where $N_{A{\rm Au}}(X)$ and $N_{\rm atom}(X)$ are the L\"{o}wdin charge obtained from the calculations for the $A$Au in the B2 structure and the atom in a supercell, respectively. Due to the small electronegativity of the alkali metals, the electron transfer occurs from the alkali metals to Au, yielding negative value of $\Delta N_{A{\rm Au}}(A)$. For $A=$ Rb, Cs, and Fr, $\Delta N_{A{\rm Au}}(A)=-0.35, -0.96$, and $-1.00$, respectively. This indicates that the ionic character of the bonding becomes strong when the Rb is replaced with the Cs atom in the $A$Au. However, the amount of the charge transfer is almost saturated at $A=$ Cs as one goes from $A=$ Rb to Fr. 

{\it Summary.---}We have performed DFT and DFPT calculations on the B2 FrAu as well as RbAu and CsAu. We have demonstrated that the FrAu has a relatively large $a$ and a relatively small $B$ and thus identified as an outlier in the B2 compounds. The FrAu in the B2 structure is predicted to be dynamically stable through the phonon dispersion calculations. Overall, the structural and electronic properties of the FrAu are quite similar to those of the CsAu. The hypothetical FrAu is included to one of the ionic compounds. 

\begin{acknowledgments}
This work was supported by JSPS KAKENHI (Grant No. 21K04628). A part of numerical calculations has been done using the supercomputer ``Flow'' at Information Technology Center, Nagoya University.
\end{acknowledgments}




\end{document}